\documentclass[final,3p,times,twocolumn]{elsarticle}

\usepackage{graphicx}
\usepackage{hyperref}
\usepackage{xcolor}
\usepackage{amsmath}
\usepackage{amsfonts}
\usepackage{amssymb}
\usepackage{lineno}
\usepackage{anyfontsize}
\usepackage{ulem}

\def\be{\begin{equation}}
\def\ee{\end{equation}}
\def\bea{\begin{eqnarray}}
\def\eea{\end{eqnarray}}

\def\bean{\begin{equation*}}
\def\eean{\end{equation*}}

\newcommand{\half}{{\frac{1}{2}}}

\newcommand{\req}[1]{(\ref{#1})}

\def\nn{\nonumber}

\newcommand{\up}{{\uparrow}}
\newcommand{\dw}{{\downarrow}}

\def\bec{\begin{center}}
\def\enc{\end{center}}
\def\lb{\label}

\def\he{$\rm ^4He ~$}

\begin{document}

\title{QCD Hidden-Color Hexadiquark in the Core of Nuclei}
%\title{QCD Hidden-Color Hexadiquark in the Core of Nuclei:  A Novel Explanation of the EMC Effect}

\author[LBL,JLAB,SLAC]{Jennifer~Rittenhouse~West}%\email{jennifer@lbl.gov}
\author[SLAC]{Stanley~J.~Brodsky}%\email{sjbth@slac.stanford.edu}
\author[UCR]{Guy~F.~de~T\'eramond}%\email{gdt@asterix.crnet.cr}
\author[SUNY]{Alfred~S.~Goldhaber}%\email{goldhab@max2.physics.sunysb.edu}
\author[UTFS]{Iv\'an~Schmidt}%\email{ischmidt@fis.utfsm.cl}

\address[LBL]{Lawrence Berkeley National Laboratory, Berkeley, CA 94720, USA}
\address[JLAB]{The EIC Center at Jefferson Laboratory, Newport News, VA 23606, USA}
\address[SLAC]{SLAC National Accelerator Laboratory, Stanford University, Stanford, CA 94309, USA}
\address[UCR]{Laboratorio de F\'isica Te\'orica y Computacional, Universidad de Costa Rica, 11501 San Jos\'e Costa Rica}
\address[SUNY]{C.N. Yang Institute for Theoretical Physics, State University of New York, Stony Brook, NY 11794-3840, USA}
\address[UTFS]{Departamento de  F\'isica y Centro Cient\'ifico Tecnol\'ogico de Valpara\'iso-CCTVal, Universidad T\'ecnica Federico Santa Mar\'ia, Casilla 110-V, Valpara\'iso, Chile}

\date{\today}

\begin{abstract}

Hidden-color configurations are a key prediction of QCD with important physical consequences. 
In this work we examine a QCD color-singlet configuration in nuclei formed by combining six scalar $[u d]$ diquarks in a strongly bound $\rm SU(3)_C$ channel. The resulting hexadiquark state is a charge-2, spin-0, baryon number-4, isospin-0, color-singlet state.  It contributes to alpha clustering in light nuclei and to the additional binding energy not saturated by ordinary nuclear forces in \he as well as the alpha-nuclei sequence of interest for nuclear astrophysics. We show that the strongly bound combination of six scalar isospin-0 $[ud]$ diquarks within the nuclear wave function - relative to free nucleons -  provides a natural explanation of the EMC effect measured by the CLAS collaboration's comparison of nuclear parton distribution function ratios for a large range of nuclei. These experiments confirmed that the EMC effect; i.e., the distortion of quark distributions within nuclei, is dominantly identified with the dynamics of neutron-proton (``isophobic'') short-range correlations within the nuclear wave function rather than proton-proton or neutron-neutron correlations.
\vspace{1pt}

\end{abstract}

%\begin{keyword}
%\sep
%\end{keyword}

\maketitle

\section{Introduction}
\label{sec:intro}

A striking feature of the nuclear structure functions measured in deep inelastic scattering (DIS) is the strong deviation from nucleon additivity observed by the European Muon Collaboration (EMC) at CERN in the kinematic domain of the Bjorken scaling variable $0.3 < x_{Bj} < 0.7$ \cite{Aubert:1983xm,doi:10.1142/S0218301314300136}.  From a nuclear physics perspective, the EMC effect is hypothesized to be due to either mean-field nuclear effects or short-range two nucleon interactions.  In this work we take the latter view.  The EMC effect is taken to be a direct measure of strong internal nucleon-nucleon dynamical interactions related to short-range correlated (SRC) nucleon pairs within the nucleus \cite{Weinstein:2010rt} rather than a static modification of the nuclear mean field \cite{Norton:2003cb, Cloet:2019mql}.  Nuclear shadowing and anti-shadowing effects are also observed in the nuclear parton distribution functions (PDFs) at low $x_{Bj} $, but these effects reflect diffractive processes and the interference of single and multiple scattering amplitudes~\cite{Brodsky:1989qz}.  The EMC effect is observed in deep inelastic scattering DIS experiments at leading twist, meaning that the quark structure of the nuclear target is directly measured.  In this paper we analyze the dynamical EMC deviation of nuclear structure functions from nucleon additivity at a fundamental level from QCD degrees of freedom.

The CLAS collaboration recently analyzed the detailed dependence of the EMC effect on the composition of the nuclear target, systematically comparing nuclei and isotopes with different numbers of protons and neutrons, by utilizing simultaneous measurements of DIS and quasi-elastic scattering~\cite{Schmookler:2019nvf}. The physical picture that emerges from their analysis is of high virtuality nucleons in the nucleus  fluctuating  into strongly interacting SRC pairs, thus distorting their internal quark and gluon structure.  The short-range correlations appear to be specifically associated with neutron-proton scattering events versus neutron-neutron or proton-proton interactions within the nuclear domain~\cite{Schmookler:2019nvf}.  This pattern of SRCs has a remarkably strong isospin dependence in the nucleon-nucleon interaction.  As stated by the CLAS collaboration, the SRC of nucleons within nuclei appear to be `isophobic'; i.e., similar nucleons are much less likely to be correlated than dissimilar nucleons, leading to many more neutron-proton SRC pairs than neutron-neutron and proton-proton pairs~\cite{Duer:2018sby}. Indeed, the neutron-proton correlated pairs are as much as 20 times as prevalent as the proton-proton and neutron-neutron pairs~\cite{Subedi:2008zz}.\footnote{A recent reanalysis of CLAS data indicates a transition to spin-independent correlations at very high momenta likely due to  universal $N-N$ repulsion at very short distances~\cite{Schmidt:2020kcl}.}  The isospin structure of nucleon-nucleon SRC pairs was first described using Alternating Gradient Synchrotron (AGS) accelerator data from Brookhaven National Laboratory~\cite{Piasetzky:2006ai}.

\section{The hexadiquark model \lb{model}}
The hadronic (and nuclear) eigenstates of the QCD light-front (LF) Hamiltonian $H_{LF} |\Psi_H\rangle = M^2_H |\Psi_H\rangle$  are built on LF Fock states $|n\rangle,$ the color-singlet eigenstates of the free LF Hamiltonian $ H^0_{LF}$.  
The coefficients  $\langle n|H\rangle$ in the Fock state expansion for hadrons are the LF wave functions $\Psi_H(x_i, k_{\perp i}, \lambda_i)$  which depend only on the internal partonic coordinates: the longitudinal momentum fraction $x_i$, the transverse momentum $k_{\perp i}$, and the constituent spin along the $z$-direction $\lambda_i$. The LF wave functions underlie hadronic observables such as form factors,  parton distribution functions, transverse momentum dependent distribution (TMDs), fragmentation functions and distribution amplitudes \cite{Brodsky:1997de}. 
 
This LF Fock state expansion has led to novel perspectives for the nonperturbative QCD structure of hadrons including the quark-antiquark structure of mesons, the quark-diquark structure of baryons, and the diquark-antidiquark structure of tetraquarks. One such theoretical perspective is the light-front holographic QCD (LFHQCD) approach where the color-confining interaction is determined by an underlying superconformal algebraic structure \cite{Brodsky:2020ajy}.  

LFHQCD is based on the gauge/gravity correspondence between classical gravity in five dimensions and light-front quantization in the physical 3 + 1 space-time dimensions \cite{Brodsky:2014yha}.  It predicts that the three-quark valence state of the proton is in a quark-diquark configuration $|u [ud]\rangle$ where $[ud]$ is a scalar diquark with color ${\bf{\overline 3}_C}$.  LFHQCD gives a systematic accounting of observed hadron spectroscopy including the massless pion in the chiral limit. It also predicts supersymmetric 4-plet relations between the meson, baryon and tetraquark eigenstates and their Regge trajectories with universal slopes \cite{Brodsky:2020ajy}.

%Hidden-color configurations are an important prediction of QCD with measurable physical consequences. In the case of the deuteron, the I=0, J=1 valence Fock state is expanded in five color-singlet combinations  $|uuuddd\rangle$ of six up and down quarks. Only one of these Fock state corresponds to the standard two nucleon state $|np\rangle$. 
%[No! Fix!] The other ``hidden-color'' Fock states are relevant for deuteron phenomena at short distances such as the deuteron form factor at large momentum transfer or the deuteron structure function at large $x_{Bj}~$ \cite{1983PhRvL..51...83B}. 

For the EMC effect, we propose a novel QCD Fock state expansion for nuclei -- the existence of a B=4, Q=2 spin, isospin and color-singlet configuration, the hexadiquark (HdQ),
\be \lb{H}
| {\psi}_{\rm HdQ} \rangle = |[ud] [ud] [ud][ud] [ud] [ud]\rangle ,
\ee
which is composed of a strongly bound set of six scalar diquarks with the same global quantum numbers as the \he nucleus, $\rm J^{P}=0^{+}$.

Each neutron-proton pair donates three $[ud]$ diquarks to the HdQ, with the proton donating a valence $u$ and the neutron a valence $d$ to create the third diquark.  Isophobic SRCs are due to the strong QCD interactions of the $u$ and $d$ bond within each diquark of the six-diquark HdQ in the nuclear environment, not from nucleon-nucleon n-p interactions within the nucleus.  To illustrate this point, consider the case of a semi-inclusive DIS experiment on a nucleus where the distribution of two final state nucleons is measured.  The struck $u$ quark and its recoiling $d$ quark partner in any of the diquarks within the HdQ will produce outgoing nucleons $p=  |u [ud]\rangle$ and $ n= |d [ud]\rangle$ with balancing momenta.  This will appear in such an experiment as an n-p SRC.

The HdQ state is a color-singlet combination of six I=0, J=0 $[ud]$ diquarks, each in the antisymmetric color
 configuration  $\bf \overline 3_C$,  a remarkable QCD bound state of 12 quarks in a single domain of color confinement.  In addition, we shall show that the six diquarks are correlated as three ${\bf\overline 6}_C$ scalar duo-diquarks (DdQs) $[[ud][ud]]$. The HdQ state is a manifestation of  a ``hidden-color" \cite{1983PhRvL..51...83B,West:2019eti}
nuclear state since it matches the quantum numbers B=4, A=4, Q=2,  I=0, $\rm J^P=0^+$ of the $^4$He nucleus. It is thus a viable candidate for
the underlying isophobic SRC structure function distortions behind the EMC effect.

\section{Construction of the hexadiquark wave function \lb{HdQWF}}

The hexadiquark is a $\rm J^P= 0^+$, I=0, B=4, Q=2 hadron constructed in a three-step process.  In the first step, two quarks  in the fundamental color representation ${\bf 3}_C$ form a $[ud]$ scalar diquark which transforms as a $ {\bf \overline 3}_C$ color object in the product ${\bf 3}_C \otimes  {\bf 3}_C \to  {\bf6}_C  \oplus{\bf  \overline3}_C$. The ${\bf \overline 3}_C$ representation is chosen over the ${\bf6}_C$ because it is antisymmetric, and with the scalar diquark transforming as singlet under spin and space, this builds a totally antisymmetric nucleon wave function in compliance with Fermi statistics. In the second step, two scalar diquarks in the ${\overline 3}_C$ combine to form a duo-diquark (DdQ) which transforms in the symmetric color representation ${\bf \overline {6}}_C$ in the product ${\bf \overline 3}_C \otimes  {\bf \overline  3}_C \to {\bf \overline 6}_C \oplus {\bf 3}_C $ \req{DTP}. It is not possible to form a DdQ from two identical scalar diquarks in a relative $S$-state in the  antisymmetric  ${\bf 3}_C$ because it vanishes identically. On the other hand, it is possible to construct a DdQ in the ${\bf \overline 6}_C$ which is also  a singlet under isospin and space rotations: Its total wave function is totally symmetric according to the spin-statistics theorem. For the third and final step, the three DdQs bind together to form the color singlet HdQ in the $\rm SU(3)$ product  ${\bf \overline 6}_C \otimes   {\bf \overline 6}_C \otimes {\bf \overline 6}_C \to {\bf 1}$.  This is viable because ${\bf \overline 6}_C \otimes   {\bf \overline 6}_C \to {\bf  6}_C$, and the product  ${\bf 6}_C \otimes   {\bf \overline 6}_C$ contains the color singlet:
\be \lb{3bar6DdQ}
\left\{\left([u d]_{{\bf \overline 3}_C} [u d]_{{\bf \overline 3}_C}\right)_{{\bf \overline 6}_C}\! \left([u d]_{{\bf \overline 3}_C} [u d]_{{\bf \overline 3}_C}\right)_{{\bf \overline 6}_C} \! 
\left([u d]_{{\bf \overline 3}_C} [u d]_{{\bf \overline 3}_C}\right)_{{\bf \overline 6}_C} \right\}_{{\bf \overline 1}_C}.
\ee
This product of $\bf \overline 6_C$ duo-diquark clusters has identical quantum numbers as the $\alpha$ particle and overlaps with the usual QCD nuclear bound state 
\be 
\left\{{\left(u_{{\bf 3}_C}[ud]_{{\bf \overline 3}_C}\right)_{{\bf 1}_C} \!\left(d_{{\bf 3}_C}[ud]_{{\bf \overline 3}_C}\right)_{{\bf 1}_C} \!\left(u_{{\bf 3}_C}[ud]_{{\bf \overline 3}_C}\right)_{{\bf 1}_C} \!\left(d_{{\bf 3}_C}[ud]_{{\bf \overline 3}_C}\right)_{{\bf 1}_C}}\right\}_{{\bf 1}_C},
\ee
expressed in terms of identical quarks but clustered into nucleons.  Isoscalar diquarks in the ${\bf 6}_C$ have been used to build recently detected hexaquark wave functions~\cite{Kim:2020rwn}.

Computing the global quantum numbers and examining the full configuration for this 12-body bound-state system is a complex dynamical problem.  The discussion is greatly simplified when considered in terms of the DdQ effective bosonic degrees of freedom, neglecting additional relative coordinates, for example, from the quarks in a diquark or the excitation between diquarks in a DdQ.  This reduces the number of independent variables and degrees of freedom to a minimum.  Following this procedure, the 12 quarks are organized into 3 DdQ clusters in the ${\bf \overline 6}_C$ representation of SU(3) as shown in Eq.\req{3bar6DdQ}.

\begin{table}[htp]
\caption{Effective SU(3) color factors $C_F$, Eqs. \req{CasF} and \req{CasFn}, in various formation channels from one-gluon exchange: The minus (plus) sign in the third column corresponds to short-range attraction (repulsion).  The label C or NC refers to spin-statistics compliant or non-compliant cluster configurations.}
\begin{center}
\begin{tabular}{|c|c|c|c|}
\hline\hline
Configuration & Channel &   $C_F$  & C/NC \\
\hline
Diquark &  $\bf 3 \otimes  \bf 3 \to \bf \overline 3$  &  - 2/3   & C \\
             &  $\bf 3 \otimes  \bf 3 \to \bf   6$           &  ~ 1/3   & C \\
\hline
DdQ &  $\bf \overline3  \otimes  \bf \overline 3 \to \bf 3$  &  - 2/3   & NC\\ 
        &  $\bf \overline3  \otimes  \bf \overline 3 \to \bf \overline 6$  &  ~1/3   & C\\  
\hline
 2 DdQ &  $\bf \overline 6  \otimes  \bf \overline 6 \to \bf 6$  &  - 5/3   & C\\    
 \hline
 HdQ &  $\bf \overline6  \otimes  \bf \overline 6   \otimes  \bf \overline 6 \to \bf 1$  &  - 5   & C\\         
\hline \hline
\end{tabular}
\end{center}
\label{CF}
\end{table}

One-gluon exchange is attractive in the  ${\bf3}_C \otimes {\bf 3}_C \to {\bf\overline 3}_C$ diquark  channel; in contrast, the short-range interaction in the ${\bf3}_C \otimes {\bf 3}_C \to  {\bf6}_C$ channel is repulsive (Table \ref{CF}). Likewise, the allowed DdQ formation channel ${\bf \overline 3}_C \otimes {\bf \overline 3}_C \to  {\bf \overline 6}_C $ is repulsive at short distances, but as a counter to this repulsion the DdQ will remain color confined at larger distances at a radius determined by the QCD scale. Finally, one-gluon exchange between two HdQs in the clustering channel ${\bf \overline 6}_C \otimes {\bf \overline 6}_C \to  {\bf 6}_C$ leads to short-range attraction. The HdQ interactions in the color-singlet HdQ channel ${\bf \overline 6}_C \otimes {\bf \overline 6}_C \otimes  {\bf \overline 6}_C \to {\bf 1}$ leads to a strong short-range attraction simply by adding color factors for every two-body interaction beginning with the diquark.  The color factor for this construction is $C_F = -5$, as compared with the corresponding factor for the  nucleon, $C_F = -2$, in its color singlet representation.

The short-range repulsion in the second step of the construction, with two diquarks forming a ${\bf \overline 6}_C$ of SU(3), should lead to a more extended object, probably having a larger overlap with the usual nuclear bound state. In fact, the HdQ hidden-color configuration may be particularly relevant at large distances where the HdQ should contribute to alpha clustering in light nuclei~\cite{Wheeler:1937zza, Fossez:2020pqu} and to the additional binding energy observed in the alpha-nuclei sequence, e.g., in the ground states for $^4{\rm He}, ^8\!{\rm Be}, ^{12}\!{\rm C}, ^{16}\!{\rm O}, \dots$, and states not saturated by the standard nuclear forces.

 To construct the HdQ wave function we follow the three-step procedure described above.
The scalar diquark $\psi^{[ud]}_a$ is given by the spin-isospin singlet product 
\bea \lb{sdq}
\psi^{[ud]}_a &=& [ u d ]_a  \\
 &= &\frac{1}{\sqrt 2}  \epsilon_{a b c}\big( u^b\up d^c \dw   - d^b\up u^c\dw \big) \nn,
\eea
where the indices $a, b, c = 1, 2, 3$ are color indices in the fundamental $\rm SU(3)_C$  representation. The scalar diquark is a $\rm J^P= 0^+, I =0$ object which transforms as color ${\bf \overline 3}$.

In the second step we construct the DdQ, $\psi^{[udud]}$,  the product  ${\bf \overline 3}_C \otimes  {\bf \overline  3}_C$ from two scalar diquarks.  It is the sum of a ${\bf 3}_C$ and  a ${\bf \overline 6}_C$ represented by the  symmetric tensor \req{psibar6}. The DdQ, $\psi^{[udud]}$, is thus given by the symmetric tensor operator
\be \label{DdQ}
\psi_{a b}^{[udud]} = \psi^{[ud]}_a \psi^{[ud]}_b,
\ee
an isospin singlet state which transforms in the symmetric ${\bf \overline 6}$ color representation under $\rm SU(3)_C$ transformations.  The DdQ itself is also an effective scalar boson since it is the product of two scalar bosons: It transforms as a $\rm J^P = 0^+$ state under SO(3) rotations.

 Lastly, we construct the HdQ which is the color singlet product of three DdQ in the ${\bf \overline 6}_C$. To this end, we first construct the symmetric ${\bf 6}_C$  out of the product  of two  ${\bf \overline 6}_C$ in the complex conjugate representation: ${\bf \overline 6}_C \otimes   {\bf \overline 6}_C  \to {\bf  6}_C$. It is given by the symmetric tensor \req{psi6}.
The HdQ wave function, $\psi_{HdQ}$, is thus the color singlet 
\be \label{HdQ}
\psi_{HdQ} = \epsilon^{a c f } \epsilon^{b d g}  \,  \psi_{a b}^{[udud]}  \psi_{c d}^{[udud]}  \psi_{f g}^{[udud]},
\ee
 which, as required, is fully symmetric with respect to the interchange of any two bosonic duo-diquarks. The HdQ spatial wave function must also be totally symmetric with respect to the exchange of any two DdQs in order for the total wavefunction to obey the correct statistics. The HdQ is a $\rm J^P = 0^+$, I=0 color singlet state,  matching the  quantum numbers of the \he nucleus ground state.

The physical alpha particle eigenstate of the QCD Hamiltonian is a linear combination of both the hidden-color HdQ and the conventional four-nucleon bound state of nuclear theory
\begin{multline}
|\alpha\rangle = C_{pnpn} \left|(u[ud])_{{\bf 1}_C} (d[ud])_{{\bf 1}_C} (u[ud])_{{\bf 1}_C} (d[ud])_{{\bf 1}_C} \right\rangle \\
+ ~~C_{\rm HdQ} \left|([ud][ud])_{{\bf \overline 6}_C}([ud][ud])_{{\bf \overline 6}_C}([ud][ud])_{{\bf \overline 6}_C}\right\rangle.
\end{multline}

The mixing of the HdQ with the conventional $|npnp\rangle$ nuclear state will also lead to a new $\rm J^P=0^+$ excited eigenstate which can decay to the ground state by virtual photon emission.  The presence of the HdQ hidden-color Fock state, which is strongly bound by the QCD color-confining interactions, can lower the mass of the $^4$He nuclear eigenstate, and thus can account for its exceptionally strong $\sim 7$ MeV binding energy per nucleon. Nuclei with $A=2Z \ge 4$, which can be identified as clustered multi-alpha  bound states such as $^{12}$C and $^{16}$O,  have increased binding relative to their neighbours in the nuclear isotope sequences and can have multi-HdQ components. In principle, other heavy nuclei with $A > 4$ can have both multi-HdQ and nucleon degrees of freedom. 

One might expect nuclear HdQ effects to be small as nuclear computations based on the nucleon-nucleon interaction and chiral effective field theory give a reasonable description of binding energies for light nucleons. In fact, recent lattice simulations using effective nucleon degrees of freedom to compute the ground state properties for $\sim 70$ isotopes from $^3$He to $^{48}$Ca, give binding energies and charge radii with a typical NLO error of 10\%~\cite{Lu:2018bat}.  However, the physics contained within the free parameters allows for the possibility of larger effects.

It is interesting to note that the local scalar four-nucleon interaction, which obeys Wigner's approximate spin-flavor SU(4) symmetry~\cite{Wigner:1936dx},  can arise from the  HdQ  hadronic state mixing with  four nucleons:   $(p^\uparrow p^\downarrow n^\uparrow n^\downarrow)$:   All quantum numbers are identical: Q=2, B=4, I=0, J=0.  Thus the  four-nucleon  $\rm SU(4)$ scalar interaction term in the effective nuclear theory expansion~\cite{Lu:2018bat}  can be related in QCD  to the mixing of nucleonic degrees of freedom with HdQs.  This is particularly relevant to the study of alpha-like  configurations  across the light, medium and heavy mass nuclei.

The actual quantitative determination of the HdQ probability, $\mathcal{P}_{\rm HdQ} = |C_{\rm HdQ}|^2$, in nuclei with $A \ge 4$  is an important nonperturbative QCD question. Its precise value is a dynamical question and could, in principle, be answered by lattice gauge theory (LGTH), just as LGTH has recently established the importance of nonperturbative intrinsic heavy quark Fock states in hadrons \cite{Sufian:2020coz}.  The maximum contribution of the HdQ to the binding energy of $^4$He approaches the full 7 MeV per nucleon. There is a binding energy increase from the usual dependence on the atomic number $A$, which is proportional to the number of nucleon pairs and thus roughly proportional to $A^2$ for light nuclei below the saturation point; this implies the  binding energy per nucleon should grow at least as $A$~\cite{Heisenberg:1932dw}. There is also a contribution to the binding energy from closing the neutron and proton shells in $^4$He.  The maximum contribution from the HdQ state may be less than 7 MeV per nucleon.  

It is important to notice that in the nuclear shell model the nucleons move almost freely in an average nuclear potential, and the shell structure is determined mainly by the exclusion principle.  In contrast, the quarks and diquarks in the HdQ are strongly correlated in the extended confinement domain determined by the color interactions of the six ${{\bf \overline 3}_C}$ diquarks and three ${\bf 6}_C$ duo-diquarks as we have discussed above: The confining HdQ potential includes three- and four-body forces, due to gluon three- and four-vertex interactions.  In total, the HdQ state has 12 quarks with a different quantum number labeled by 3 colors, 2 spins, and 2 isospins, the maximum number in a single spatial domain allowed by spin-statistics.  In this sense, the HdQ has a `shell’ structure that is also determined by the exclusion principle, but at the quark level.

\section{Hexadiquark Configurations in Nuclei and the EMC Effect}

The EMC effect is observed as the difference between the quark PDF distribution  $q_A(x,Q)$ measured in DIS of a lepton beam of energy $Q$ on a nucleus vs. the sum of quark distributions obtained from the DIS scattering on the corresponding free nucleons:
\be \label{qA}
\Delta q_A  = q_A(x,Q) - [Z q_p(x,Q) + (A-Z) q_n(x,Q)].
\ee
In this picture, the valence structure of the sum of free nucleons corresponds to $A$ diquarks  plus $Z$ up quarks and $A-Z$ down quarks.  
However, when the nucleons are in the nuclear bound state, these $u$ and $d$ quarks are attracted to each other by the QCD color interaction and can form additional $[ud]$ diquarks.  A set of six diquarks can combine into the color singlet HdQ.

We postulate that all nuclei with $A \geq 4$ have an underlying substructure  containing one or more strongly bound hexadiquarks. The EMC effect in DIS then arises from the lepton scattering on an up or down quark in an I=0, S=0, Q= $+\frac{1}{3}$ $[ud]$ diquark within the Q=2, B=4, I=0 hexadiquark state. The struck up or down quark is strongly correlated with a quark of opposite isospin. The correlation therefore appears as an isophobic short-range correlation.

More precisely, suppose an $u$ quark struck by the lepton in DIS is in one of the six $[ud]$ diquarks. Its light-front longitudinal momentum fraction $x$-distribution is modified by the strong interactions with its $d$-quark partner. Similarly, if the lepton scatters on a $d$ quark in an $[ud]$ diquark, its  LF $x$-distribution is modified by the strong interactions with its $u$-quark partner.   The pairing of valence up and down quarks in conjunction with existing diquarks creates isophobic SRCs at the quark level.

The difference of the quark distribution $\Delta q_A $  in nuclei vs. free nucleons measured by EMC and CLAS  can be attributed to DIS on the quarks of the four extra diquarks formed from 2 sets of nearest neighbor N-N interactions in the nuclear eigenstate.  Moreover the N-N SRCs arising from the existence of the set of six scalar diquarks are automatically isophobic.

\section{Additional Consequences of the Hexadiquark Model}
The HdQ component within the $^4\rm He$ nuclear wavefunction may play a crucial role in lowering its mass.  Diquarks can form through attractive SU(3)$_C$ interactions and diquarks themselves can bind together leading to the hexadiquark with the spin-color  and clustering structure of Eq.\req{3bar6DdQ}. It is the twelve quark color singlet manifestation of hidden color \cite{1983PhRvL..51...83B} in the nuclear system.

Several questions are raised by the scalar hexadiquark model.  

1.  Does this explanation for the EMC effect imply it is identical for $u$ and $d$ type quarks?

The effect is identical for $u$ and $d$ type quarks.  We have not given a special role to the specific $u$ and $d$ coupling to the virtual photon in the DIS sequence for the formation of the intermediate HdQ state.  Therefore the basic mechanism would not depend on the $u$ or $d$ active quark; however the specific composition of the debris in the final state would depend on the $u$ or $d$ coupling. Nuclei with $Z<\frac{A}{2}$, i.e., those with more neutrons than protons as is the case for the majority, will therefore appear to interact more frequently with up quarks.  However the HdQ treats both $u$ and $d$ on the same footing. 

2.  Does the model imply that there is an EMC effect for sea quarks?  

A through exploration of hexadiquark dynamics would be required to answer this question  precisely. As discussed earlier, we do expect larger repulsive forces at the core from $\overline{{\bf 6}}$ duo-diquarks, which would exceed the larger distance confinement pressure from gluons \cite{Burkert:2018bqq}.  Therefore the effects from higher Fock components would likely be decoupled as compared to the valence contribution of the intermediate hexadiquark. 

3.  Does the EMC effect depend on the target polarization?

The model proposed is isospin dependent and in fact is isophobic. Therefore even for n-p pairs there is an important dependence on the singlet or triplet configuration of the nucleon pair. One could devise a polarization experiment which could address this critical prediction of the model.

4.  Is the EMC effect the same for DIS charge current; i.e., neutrino beam  vs. neutral current reactions?

The equivalent DIS sequence for the charged current amounts to replacing the $\gamma*$ virtual photon by a spin-1 $W$ vector boson.  Since the $W$ has no isospin assignment, we expect processes similar to the sequence initiated by an isospin-0, J=1 photon.

\section{Conclusions and Outlook}
\label{sec:conclu}
The existence of a color singlet hexadiquark configuration in nuclear QCD eigenstates may have a major impact on fundamental nuclear structure theory. The $\rm SU(3)_C$ interactions in the nuclear environment allow the formation of a strongly bound isospin-singlet sextet of scalar diquarks in the attractive channel as a QCD state of the nuclear system.  In this picture, the quark-diquark model of individual nucleons is modified in the nuclear wave function by nearest neighbor interactions creating a set of six diquarks between four nucleons that form the HdQ configuration.

Furthermore, we have identified a mechanism based on QCD degrees of freedom that describes the strong isospin dependence of short-range N-N correlations.  The key point for nuclear structure is the special role of the color singlet hexadiquark. It consists of six isospin-singlet scalar diquark structures strongly bound into a hidden-color QCD state in the nuclear system.  In this picture, the quark-diquark model of individual nucleons is modified during nearest neighbor short-range interactions within the nucleus.  Scalar diquarks from neighboring nucleons form through QCD interactions with valence up and down quarks, with each nucleon donating one valence quark to their shared diquark.  The flavor composition for scalar diquarks requires opposite isospin in order to satisfy the spin-statistics theorem for $[ud]$.  The hexadiquark is formed through interactions occurring between neutrons and protons rather than neutron-neutron or proton-proton (the definition of ``isophobic''). Thus the HdQ may underlie the phenomena of isophobic SRCs. 
It has zero internal orbital angular momentum,   its constituents are consistent with Fermi statistics and its quantum numbers are identical to the $\alpha$ particle. Indeed, hexadiquarks are proposed as an important degree of freedom within heavy nuclei by its mixing with nuclear states with $A \geq  4$.

The difference between leading twist nuclear and nucleon structure functions in Eq.~\req{qA} is sensitive to the difference of $[ud]$ diquarks in nuclei vs.~nucleons,  and thus the isophobic EMC effect is also sensitive to the HdQ correlations within the QCD nuclear eigenstates. This explanation can be tested by diffractively dissociating relativistic beams of heavy nuclei \cite{West:2019eti}. The final state should display the underlying hexadiquark composition of the nucleus.

The hexadiquark is formed from four nucleons:  two protons and two neutrons.  It requires momentum sharing between pairs of $u$ and $d$ valence quarks in nearest neighbor neutrons and protons, thus leading to the observed ejection of the proton and neutron containing the struck quark in the SRC pair.  A key aspect of the model is its isospin structure which gives a strong isospin dependence to the correlated pairs.  The MARATHON experiment at Jefferson Lab~\cite{ftos:2000qp} with $\rm ^3He$ and $\rm ^3H$ targets should not exhibit strong isophobic SRCs as the hexadiquark cannot form in such nuclei.  During the revision process, we've been informed of preliminary results indicating that the isophobic nature of SRC may deplete as predicted by the HdQ model  \cite{cite-key,cite-key2,cite-key3,osti_1577446,cite-key4}.  Predictions for diquarks in $A=3$ nuclei are contained within a companion work on N-N diquark formation \cite{West:2020tyo}.  Any nuclei with $A>3$ is predicted to contain the hexadiquark in its Fock space and will therefore lead to stronger binding and exhibit strong isophobic SRC, a further investigation of which we defer to future work \cite{West:2021fyi}.

 \subsection*{Acknowledgements}
We are grateful to Francesco Coradeschi, Robert Jaffe and   Jerry Miller for helpful comments and insights.  This work is supported in part by the Department of Energy, Contract DE--AC02--76SF00515.  I.S. is supported by ANID PIA/APOYO AFB180002 (Chile) and Fondecyt (Chile) grant 1180232.  SLAC-PUB-17525.

\appendix

\section{Tensor Product Decomposition and Casimir  Invariant Operators\lb{DTP}}

In this appendix we examine the decomposition of SU(3) tensor  products  which are particularly useful  to construct the wave function in different representations of the gauge group. We also indicate the relevant  group factors which determine the sign and strength of one-gluon exchange between particles in different SU(3)$_C$ representations~\footnote{Useful references for  group-theory model building are \cite{Slansky:1981yr,Georgi:1999wka,Fonseca:2020vke,Rosner:1980bd}.}

\subsection{SU(3) Tensor Product Decomposition}

We represent a vector component transforming in the fundamental representation of SU(3) by $\psi^a$ and its complex conjugate representation by $\psi_a$, where the  indices $a, b, \cdots, f = 1, 2, 3$ are indices in the fundamental  representation. We will specify an irreducible SU(3) representation by its tensor components $(n,m)$, where $n$ is the number of non-contracted upper indices and $m$ the number of non-contracted lower indices of the irreducible tensor representation with dimensionality $D(n,m)$
\be
D(n,m) = \half (n+1)(m + 1)(n + m + 2).
\ee

The product ${\bf \overline 3} \otimes {\bf \overline 3}$ is 
\be
 \phi_a \chi_b = \half \left(\phi_a \chi_b + \phi_b \chi_a\right) + \half \epsilon_{abc} \epsilon^{cde}\phi_d \chi_e,
\ee
the sum of a ${\bf \overline 6}$ given by the symmetric tensor 
\be \lb{psibar6}
\psi_{ab} [{\overline 6}] = \half \left(\phi_a \chi_b + \phi_b \chi_a\right),
\ee
and  a ${\bf 3}$, 
\be
\psi^a [3] = \epsilon^{a b c}\phi_b \chi_c.
\ee
Therefore the irreducible tensor decomposition  ${\bf \overline 3} \otimes  {\bf \overline  3} = {\bf \overline 6}  \oplus {\bf 3} $, or using the component notation $(n,m)$ the product ${\bf \overline 3} \otimes {\bf \overline 3}$  is given by $( 0,1) \otimes (0,1) =  (0,2) \oplus ( 1,0)$.

It is also illustrative to consider the product ${\bf 3} \otimes {\bf \overline 3}$
\be
\phi^a \chi_b =  \left(\phi^a \chi_b - \frac{1}{3} \delta^a_b  \phi^c \chi_c \right) + \frac{1}{3} \delta^a_b  \phi^c \chi_c ,
\ee
which is the sum of a rank-2 traceless tensor, the $\bf 8$,
\be
\phi^a_b[8] =  \left(\phi^a \chi_b - \frac{1}{3} \delta^a_b  \phi^c \chi_c \right),
\ee
and a singlet   $\bf 1$,
\be
 \psi[1] =  \phi^c \chi_c,
\ee
the trivial representation. Therefore  ${\bf  3} \otimes  {\bf \overline  3} =  {\bf  8}  \oplus  {\bf  1}$  or  $(1,0) \otimes (0,1) =  (1,1) \oplus (0,0)$.

As our last example we examine the product  ${\bf \overline 6} \otimes   {\bf 3}$:
\be
\begin{split}
\phi_{a b} \chi^{c}  =  \phi_{a b} \chi^{c} - \frac{1}{4} \delta_a^c \phi_{b d} \chi^d - \frac{1}{4} \delta_b^c \phi_{a d} \chi^d\\
+ \frac{1}{4} \left( \delta_a^c \delta_b^d + \delta_b^c \delta_a^d\right) \phi_{d f} \chi^f.
\end{split}
\ee
It is the the sum of the fifteen dimensional representation $\bf \overline{15}$
\be \label{15a}
\psi_{a b}^{c} [\overline{15}]  =  \phi_{a b} \chi^{c} - \frac{1}{4} \delta_a^c \phi_{b d} \chi^d - \frac{1}{4} \delta_b^c \phi_{a d} \chi^d,
\ee
and a $\bf \overline 3$
\be
\psi_a [\overline 3]= \phi_{ab} \chi^b.
\ee
Thus  ${\bf \overline 6} \otimes   {\bf  3} = {\bf \overline{15}} \oplus {\bf \overline 3} $ or  $(0, 2) \otimes (1, 0) =  (1,2) \oplus ( 0,1)$. Notice that the 15 dimensional representation  \req{15a} is symmetric in the lower indices and traceless: $\psi_{a c}^{c} = \psi_{c b}^{c} = 0$.

Finally, we consider the product   ${\bf \overline 6}  \otimes  {\bf \overline 6}$ given by the  \mbox{tensor} product $\phi_{ab} \chi_{cd}$ symmetric in the indices $a, b$ and $c, d$.  It is the product $(0, 2) \otimes (0,2) =  (0,4) \oplus (1, 2) \oplus (2,0)$ which corresponds to   ${\bf \overline 6} \otimes   {\bf  \overline 6} = {\bf \overline{15}’} \oplus {\bf \overline{15}} \oplus {\bf 6}$. The product ${\bf \overline 6}  \otimes  {\bf \overline 6}$ is thus decomposed into a fully symmetric  \mbox{rank-4} tensor $\psi_{a b c d}[\overline{15}]$, a rank-3 mixed tensor  $\psi_{a b}^c[\overline{15}]$ and a rank-2 symmetric tensor $\psi^{ab} [{6}]$.
We are interested in the  ${\bf 6}$: It is given by the symmetric tensor
\be \lb{psi6}
\psi^{a b}[6] = \half \left(\epsilon^{a c f} \epsilon^{b d g}  + \epsilon^{b c f} \epsilon^{a  d g} \right)  \phi_{c d} \chi_{f g} ,
\ee
which contracted with two $\bf \overline 6$ leads to a singlet from the product $ {\bf \overline 6} \otimes  {\bf \overline 6} \otimes  {\bf \overline 6} \to   {\bf 6} \otimes  {\bf \overline 6} \to {\bf 1}$. Equivalently,
\be
(0, 2) \otimes (0, 2) \otimes (0, 2) \to  (2, 0) \otimes (0, 2) \to (0,0),
\ee
the SU(3) singlet.

\subsection{Effective Color Charges}

 We consider the color interaction of particles $A$ and $B$ from the product of color charges  $T^\lambda_A$ and  $T^\lambda_B$,  where the indices $\lambda = 1, 2, \cdots, 8$ are indices in the adjoint representation of SU(3). The effective color interaction from one-gluon exchange between particles $A$ and $B$ is given by the SU(3) invariant $T_A^\lambda T_B^\lambda$,
\be
T_A^\lambda T_B^\lambda = \half \left( T^\lambda T^\lambda  - T_A^\lambda T_A ^\lambda  - T_B^\lambda T_B^\lambda \right),
\ee
where $T^\lambda = T^\lambda_A + T^\lambda_B$. If particle $A$ is in the color representation $R_A$ and particle $B$ in $R_B$, the product  $T_A^\lambda T_B^\lambda$ in the combined representation $R = R_A \otimes R_B$, 
\be
C_F \equiv T_A^\lambda T_B^\lambda \Big\vert_{R_A \otimes R_B},
\ee
is
\be \lb{CasF}
C_F = \half \left(C(R)  - C(R_A) - C(R_B) \right),
\ee
where  $C(r)$, the Casimir operator in a representation $r$, is
\be
T^\lambda T^\lambda \big \vert_r  = C(r).
\ee

For example, if particles $A$ and $B$ are in the $\bf 3$, as the two quarks composing a diquark, the relevant Casimir operators in the  $\bf 3 \otimes \bf 3 \to \bf \overline 3$ and  $\bf  3 \otimes \bf 3 \to \bf  6$ channels are
$C({\bf 3}) =  C({\bf \overline 3}) = \frac{4}{3}$, and $C({\bf 6}) =   \frac{10}{3}$ respectively. The corresponding effective color charges from \req{CasF} are $C_F = - \frac{2}{3}$ and $\frac{1}{3}$. If particles $A$ and $B$ are scalar diquarks in the $\bf \overline 3$, the only available duo-diquark (DdQ) channel is the  $\bf \overline 3 \otimes \bf \overline 3 \to \bf \overline 6$ (Section~\ref{HdQWF}).  Since $C({\bf \overline 6}) = C({\bf  6}) = \frac{10}{3}$ it follows from  \req{CasF} that $C_F = \frac{1}{3}$. Likewise, the color factor for two DdQ in the $\bf \overline 6 \otimes \bf \overline 6 \to \bf 6$ channel is $C_F = - \frac{5}{3}$, leading to short-range attraction \cite{Rosner:1980bd}.

In closing, we compute the effective color factor for an $n$-body configuration in the approximation where the interaction is the pairwise sum of two-body interactions. It is given by
\be \lb{sumT}
\sum_{i \ne j} {T_A}_i^\lambda {T_A}_j^\lambda = \half \left(T^\lambda T^\lambda  - \sum_i {T_A}_i^\lambda {T_A}_i^\lambda\right),
\ee
where $T^\lambda = \sum_i T_i^\lambda$. We compute the operator \req{sumT} in  the combined representation 
$R = R_1 \otimes R_2 \otimes \cdots \otimes R_n $ 
\be
{C_F}_n \equiv \sum_{i \ne j} {T_A}_i^\lambda {T_A}_j^\lambda  
\Big\vert_{R_1 \otimes R_2 \otimes \cdots \otimes R_n},
\ee
thus finding
\be \lb{CasFn}
{C_F}_n = \half \left(C(R)  - \sum_{i = 1}^n C({R_A}_i) \right).
\ee

Since the Casimir operator for the singlet vanishes in any representation, C(1) = 0, the total short-range color factor for the Hexadiquark (HdQ) has the value  ${\mbox C_F = -5}$ from \req{CasFn}. Notice that this lowest-order computation does not include three-gluon vertex interactions. The SU(3) color factors in the relevant channels used to build the HdQ in Section \ref{HdQWF} are given in Table \req{CF}.

%\section*{References}

\bibliographystyle{elsarticle-num}

\bibliography{hdq.bib}

\end{document}